\documentstyle[12pt,WSPRL]{article}

\begin{document}
\centerline{\normalsize\bf }
\baselineskip=22pt
\centerline{\normalsize\bf Quark Model Calculations of Symmetry}
\baselineskip=16pt
\centerline{\normalsize\bf Breaking in Parton Distributions}

\centerline{\footnotesize C. J. Benesh}
\baselineskip=13pt
\centerline{\footnotesize\it Theoretical Division, Los Alamos National
Laboratory}
\baselineskip=12pt
\centerline{\footnotesize\it  Los Alamos, NM 87545, USA}
\centerline{\footnotesize E-mail: benesh@t5.lanl.gov}
\vspace*{0.3cm}
\centerline{\footnotesize and}
\vspace*{0.3cm}
\centerline{\footnotesize T. Goldman,}
\baselineskip=13pt
\centerline{\footnotesize\it  Theoretical Division, Los Alamos National
Laboratory}
\vspace*{0.3cm}
\centerline{\footnotesize and}
\vspace*{0.3cm}
\centerline{\footnotesize G. J. Stephenson Jr.,}
\baselineskip=13pt
\centerline{\footnotesize\it Department of Physics \& Astronomy,
 University of New Mexico}
\centerline{\footnotesize\it Albuquerque, New Mexico 87131}

\vspace*{0.9cm}
\abstracts{ Using a quark model, we calculate symmetry breaking effects
in the valence quark distributions of the nucleon. In particular, we
examine the breaking of the quark model SU(4) symmetry by color magnetic
effects, and find that color magnetism provides an explanation for
deviation of the ratio $d_V(x)/u_V(x)$ from $1/2$. Additionally, we
calculate the effect of charge symmetry breaking in the valence quark
distributions of the proton and neutron and find, in contrast to other
authors, that the effect is too small to be seen experimentally.}

\normalsize\baselineskip=15pt
\setcounter{footnote}{0}
\renewcommand{\thefootnote}{\alph{footnote}}
\section{ Introduction}

	Symmetries have traditionally played a central role in our
understanding of hadrons\cite{1}. When the symmetry is
unbroken, we use it to make predictions without reference to
any model for the underlying wavefunctions. Better still, when the
symmetry {\it is} broken we can often use it as a
filter with which to study the wavefunction itself, and thus
are provided with a sensitive probe of the underlying dynamics.

	In this talk, we shall operate in the second of these modes
by examining the effect of symmetry breaking on the valence quark distributions
of the nucleon. To begin, we give a brief description of the
rationale and method used to relate the phenomenological wavefunctions
of a quark model to the parton distributions measured in high energy scattering
experiments. The following section describes the application of our
method to breaking of the quark model SU(4) spin-isospin symmetry by
color magnetic interactions\cite{JF}, and how this symmetry breaking manifests
itself in the well known difference between the $u$ and $d$ valence quark
distributions in the nucleon. Finally, we look at the case of
charge symmetry breaking by quark mass differences and by
electromagnetic effects. This effect has not yet been looked for
experimentally, but may play a small role
in the determination of $\sin^2\theta_W$ in $\nu$-nucleon
scattering\cite{2}. Although it has been suggested by some authors\cite{3}
that (relatively) large effects are to be expected, we do not find
this to be the case.

\section{ Quark Model Valence Distributions}

	We begin with a statement of the rationale that allows us to
use quark models in the study of parton distributions. Clearly, any
such attempt cannot consist of a simple evaluation of the relevant
matrix elements in terms of quark model wavefunctions, since the only
degrees of freedom in those models are the valence quarks and
(sometimes) a phenomenological representation of the confining interaction.
This picture clashes miserably with the diverse parton distributions
required by high energy
experiments, which receive large contributions from
both gluons and sea quarks. How can these two very different pictures
of a hadron be reconciled?

A possible answer lies in the
renormalization group approach of Jaffe and Ross\cite{4}. They argue
that at large momentum scales, a hadron is, as the data indicates, a
very complicated object. But as the renormalization scale is decreased,
most or all of the glue/sea found in the hadron is reabsorbed into
the valence quarks until, at very low momentum scales, the picture
changes to one in which only a relatively few degrees of freedom
are required to describe the hadron. It is this simplified picture
which one may reasonably hope to represent by a quark model.

	In the calculations described here, we shall adopt this
prescription, and proceed by evaluating the twist two contribution
to the quark distributions using quark model wavefunctions. The
resulting distributions will then be interpreted as the twist two
contribution to $q_V(x)$ evaluated at a very low renormalization
scale $\mu_{bag}^2$, and next to leading order QCD perturbation theory\cite{5}
will be used to evolve the distributions to high $Q^2$, where they can be
compared to experiment.

The matrix elements that determine the shape of the valence quark
distributions are given by
\begin{eqnarray}
q(x) &=& {1\over 4\pi} \int\, d\xi^-\,
e^{iq^+\xi^-}<N\vert\bar\psi(\xi^-)\gamma^+\psi(0)\vert
N>\vert_{LC}\nonumber\\
\bar q(x)&=& -{1\over 4\pi} \int\, d\xi^-\, e^{iq^+\xi^-}
<N\vert\bar\psi(0)\gamma^+\psi(\xi^-)\vert N>\vert_{LC}
\end{eqnarray}
where $N$ denotes the nucleon wavefunction, $q^+=-Mx/\sqrt(2)$ with
$x$ the Bjorken scaling variable, and
$LC$ denotes the light cone condition on $\xi$, $\xi^+=\vec\xi_\perp=0$.
The procedure we use is a relatively
straightforward evaluation of the matrix elements in a Peierls-Yoccoz\cite{6}
projected momentum eigenstate, the details of which may be found in
Refs. 8-10. Alternatives to this procedure may be found
in Ref. 11.

\begin{figure}

\vspace*{13pt}

\leftline{\hfill\vbox{\hrule width 5cm height0.001pt}\hfill}

\vspace*{1.4truein}             
\includegraphics{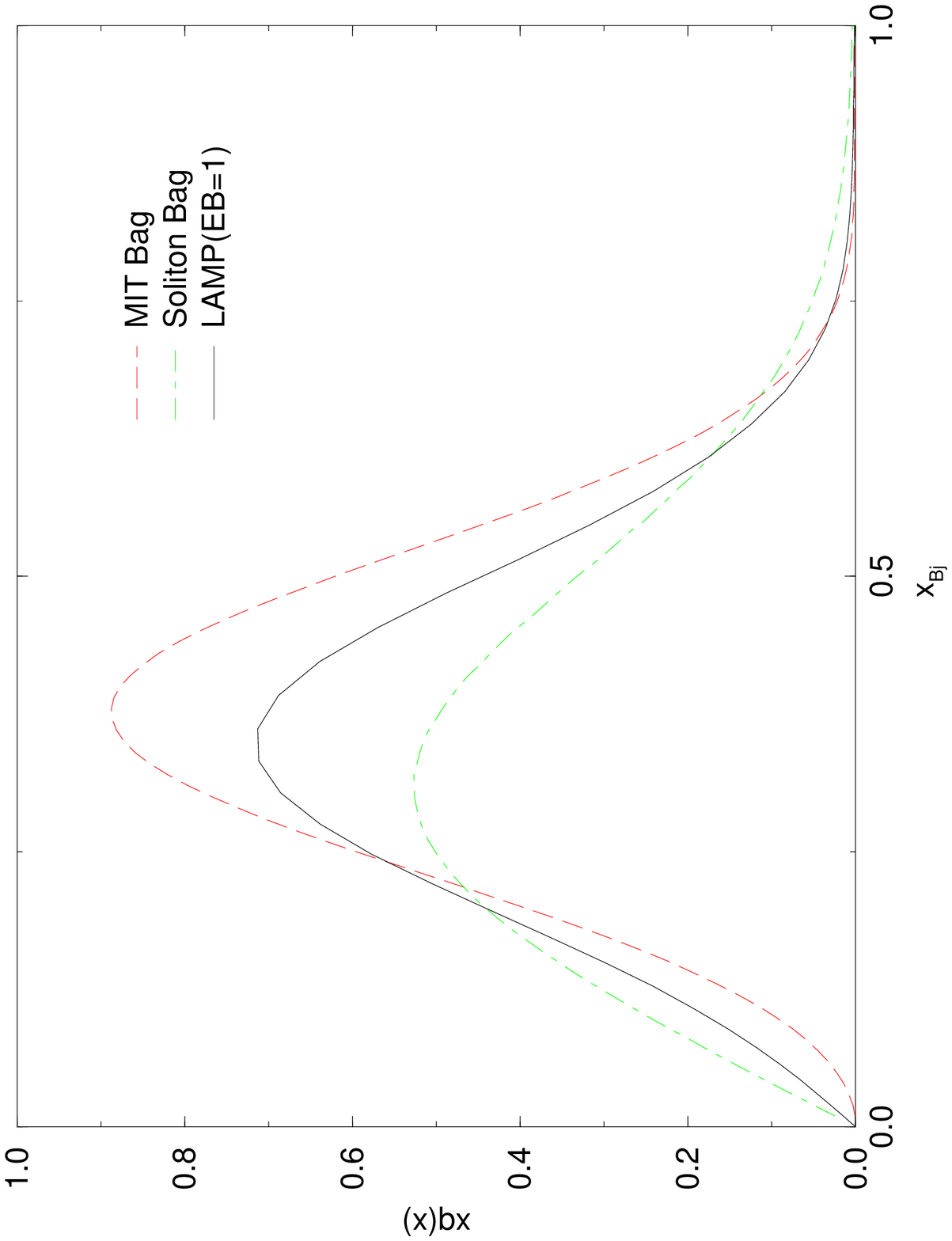}
\leftline{\hfill\vbox{\hrule width 5cm height0.001pt}\hfill}

\fcaption{SU(4) Symmetric Valence Quark Distributions at the bag scale.}

\label{fig:bagscale}

\end{figure}

	In Fig. 1, the SU(4) symmetric contribution the bag scale
valence quark distributions in the nucleon are shown for the MIT Bag\cite{6},
the Los Alamos Potential Model\cite{7}, and the soliton bag\cite{8}.
The most striking difference between these models is the area under the
graphs, which gives the momentum fraction carried by a single valence
quark. For the MIT Bag the area under the curve is nearly $1/3$,
indicative of the non-existence in the model of any degree of freedom to
 represent the
confining forces, while for the soliton bag, where the confining degree
of freedom is fully included in the calculation of the valence
wavefunctions, each valence quark carries $1/4$ of the momentum. The
LAMP model, in which confinement is implemented via a linear scalar
potential, yields a result that lies in between the others. As a
practical matter, it is advantageous to work with a model in which
a significant fraction of the hadron's momentum is carried by non-valence
degrees of freedom so that the perturbative evolution from the bag scale
to the scale at which data are taken is more trustworthy. As a
compromise, we shall carry out the symmetry breaking calculations to
come in the LAMP model, where the sharing of momentum is complemented
by the relative ease of calculation of wavefunctions.

\section{ Spin-Isospin Symmetry Breaking}

	We begin the discussion of symmetry breaking effects by
considering the SU(4) spin-isospin symmetry that appears in nearly
all quark models as a result of the spin independence of the confining
degrees of freedom and the near degeneracy of the $u$ and $d$ quark
masses. In the case of perfect SU(4) symmetry, the potential cannot
distinguish between quarks of different spin or flavor, and all
the valence quark wavefunctions and energies are the same. Hence, when
we calculate the valence quark distribution of the proton in models
with only a spin-independent interaction, the only difference between
the valence distributions $u_V(x)$ and $d_V(x)$ is the relative number
of each type of quark, and we obtain
\begin{equation}
u_V(x) = 2d_V(x)
\end{equation}
for all values of $x$. Neither this, nor the corresponding
prediction that the nucleon and $\Delta$ masses are degenerate is
manifest experimentally. In particular, the nucleon-$\Delta$ mass splitting
can be understood in terms of SU(4) symmetry breaking by either virtual
pion emission or by color magnetic interactions. In the present context,
we shall ignore the contribution due to pions, which are more likely to
contribute to sea quark distributions, and concentrate instead on the
role of color magnetism.

	Omitting the details, which may be found in Ref. 9, the
central idea is that the color magnetic interaction introduces a
dependence of the quark wavefunctions on the spin state of the other
quarks in the nucleon. Since the naive SU(4) symmetric wavefunction of
the nucleon contains correlations between the spin and flavor of
quarks required by the Pauli principle, the spin dependence of the
quark wavefunctions is transmuted into a flavor dependence of the
spin averaged wavefunctions, which in turn results in a flavor
dependence of quark momentum distributions. Neglecting color electric
effects,
the corrections to the SU(4) quark distributions can be written in
perturbation theory as
\begin{equation}
\delta q^\alpha(x) = \sum_{\alpha\neq\beta} {\bf\sigma}_\alpha\cdot
{\bf\sigma}_\beta T^{2b}(x)
+  \sum_{\alpha\neq\beta\neq\epsilon} {\bf\sigma}_\epsilon\cdot
{\bf\sigma}_\beta T^{3b}(x),
\end{equation}
where $\alpha$ denotes  the struck quark, $\beta$ and $\epsilon$ the
spectator valence quarks, and $T^{2b}(x)$ and  $T^{3b}(x)$ are functions
of the quark wavefunctions corresponding to processes in which the
gluon is exchanged between the struck quark and a spectator, and between
the spectators, respectively. As advertised, the average over the quark
spins is flavor dependent, so that
\begin{eqnarray}
\delta u(x)&=& -2T^{2b}(x)-4T^{3b}(x)\nonumber\\
\delta d(x)&=& -4T^{2b}(x)+ T^{3b}(x).
\end{eqnarray}
Physically, $T^{3b}(x)$ is an interaction between the two
spectators, so it cannot alter the {\it shape} of the struck quark's momentum
distribution. As a result of this, the ${\delta d(x)\over d(x)}$ quark is
roughly four times larger than ${\delta u(x)\over u(x)}$. Since the net
effect of the gluonic interaction is to
allow the valence quarks to lose momentum to gluons, the d quark
distribution is suppressed relative to the u quark and both
distributions carry less momentum than they would in the
SU(4) symmetric limit.

	Fixing the value of $\alpha_s$ to that required
to reproduce the nucleon-$\Delta$ mass splitting, we obtain the ratio
$d_V(x)/u_V(x)$ at the bag scale shown in Fig. 2. Also shown is the
result for the ratio at 15 GeV$^2$ , assuming $\mu_{BAG}^2=0.4$ GeV$^2$,
and $\nu$ data taken at the same scale\cite{10}.
\begin{figure}

\vspace*{13pt}

\leftline{\hfill\vbox{\hrule width 5cm height0.001pt}\hfill}

\vspace*{1.4truein}             
\includegraphics{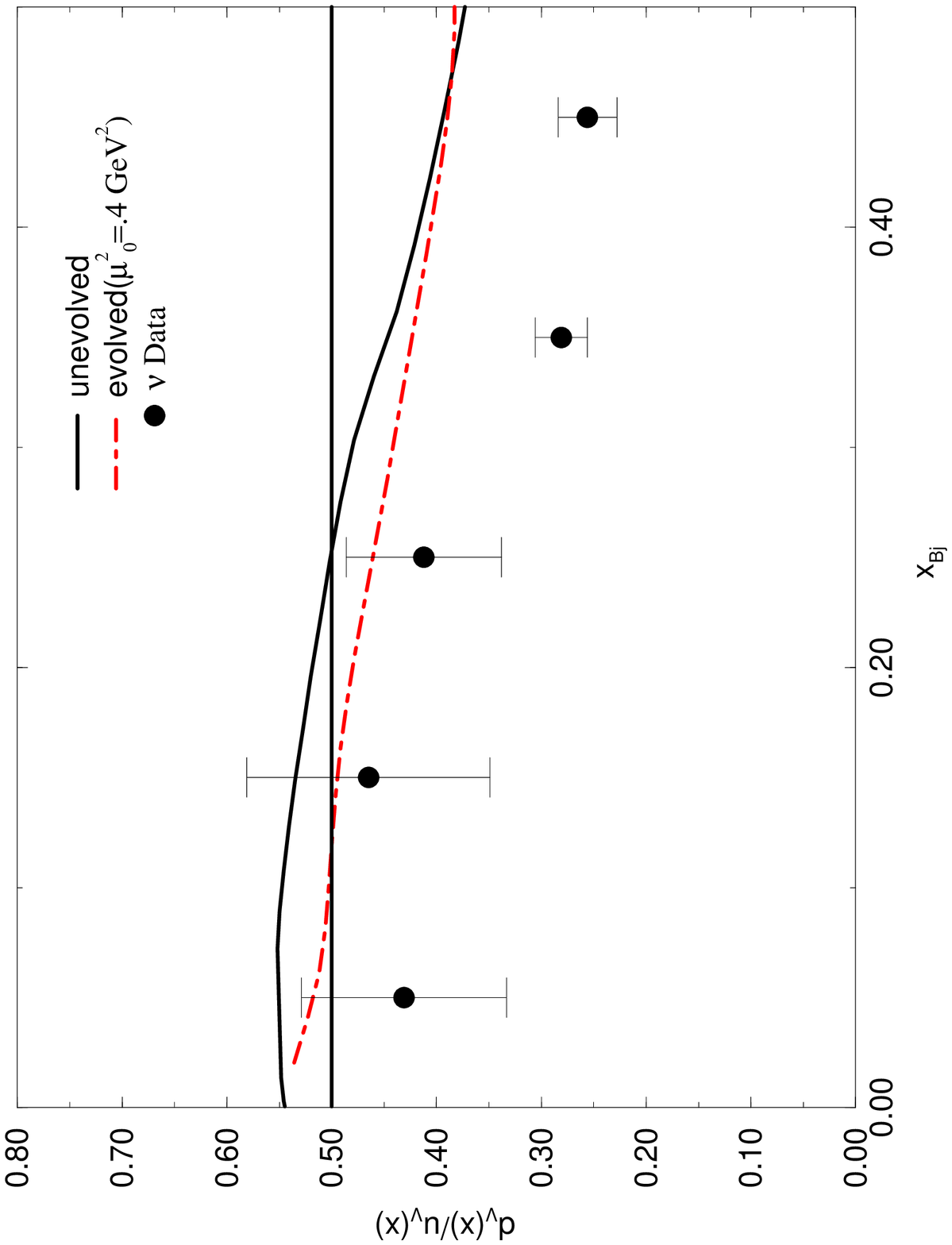}
\leftline{\hfill\vbox{\hrule width 5cm height0.001pt}\hfill}

\fcaption{ $d_V(x)/u_V(x)$ for the LAMP at the bag scale, and at 15 GeV$^2$.}

\label{fig:udratio}

\end{figure}
	While the qualitative trend( enhancement of $d_V(x)$ relative to
$u_V(x)$ at low $x$, rapidly decreasing with $x$),  is unmistakeable,
the quantitative agreement with data is less satisfactory. This
situation is expected to improve in models in which the SU(4) symmetric
valence distributions carry less momentum. In light of
this, we conclude that color magnetic  interactions provide  a natural
mechanism for producing the observed differences between the $u$ and $d$
valence distributions.

\section{ Charge Symmetry Breaking}

	Of more recent interest\cite{2} is the as yet unobserved
breaking of charge symmetry between the $u$ and $d$ quark distributions
in the nucleon. The symmetry, which is broken by $u-d$ quark mass difference
and by electromagnetic effects, is expected to be good to a few per cent.
A measure of the size of charge symmetry breaking(CSB) effects on
valence quark distributions is given by the ratios
\begin{eqnarray}
R_{min}(x) &=&2 {{d_V^p(x)-u_V^n(x)}\over {d_V^p(x)+u_V^n(x)}}\nonumber\\
R_{maj}(x) &=&2 {{u_V^p(x)-d_V^n(x)}\over {u_V^p(x)+d_V^n(x)}},
\end{eqnarray}
where $d_V^p(x)$ is the minority valence quark distribution in the proton,
$d_V^n(x)$ the majority valence quark distribution in the neutron, and so on.

We have calculated these ratios in the LAMP model\cite{11}, incorporating
a $u-d$ mass difference of 4 MeV and Coulomb effects in both the quark
energy eigenvalues and wavefunctions. In Fig. 3, we plot the result for
$R_{min}(x)$, after evolution from a bag scale of $0.4$ GeV$^2$ to 10
GeV$^2$. Also shown is the result of incorporating charge symmetry
breaking effects into the Altarelli-Parisi equation\cite{12}. For
$x<0.7$, we find that no individual contribution to CSB is greater than
about 2\%, and that for the minority quarks, the total CSB effect is
less than 4\%. For the majority quark distribution, the total CSB effect
is less than 2\%, due to partial cancellation between the effect of the
neutron-proton mass difference and the change in the quark eigenvalues.

\begin{figure}

\vspace*{17pt}

\leftline{\hfill\vbox{\hrule width 5cm height0.001pt}\hfill}

\vspace*{1.95truein}             
\includegraphics{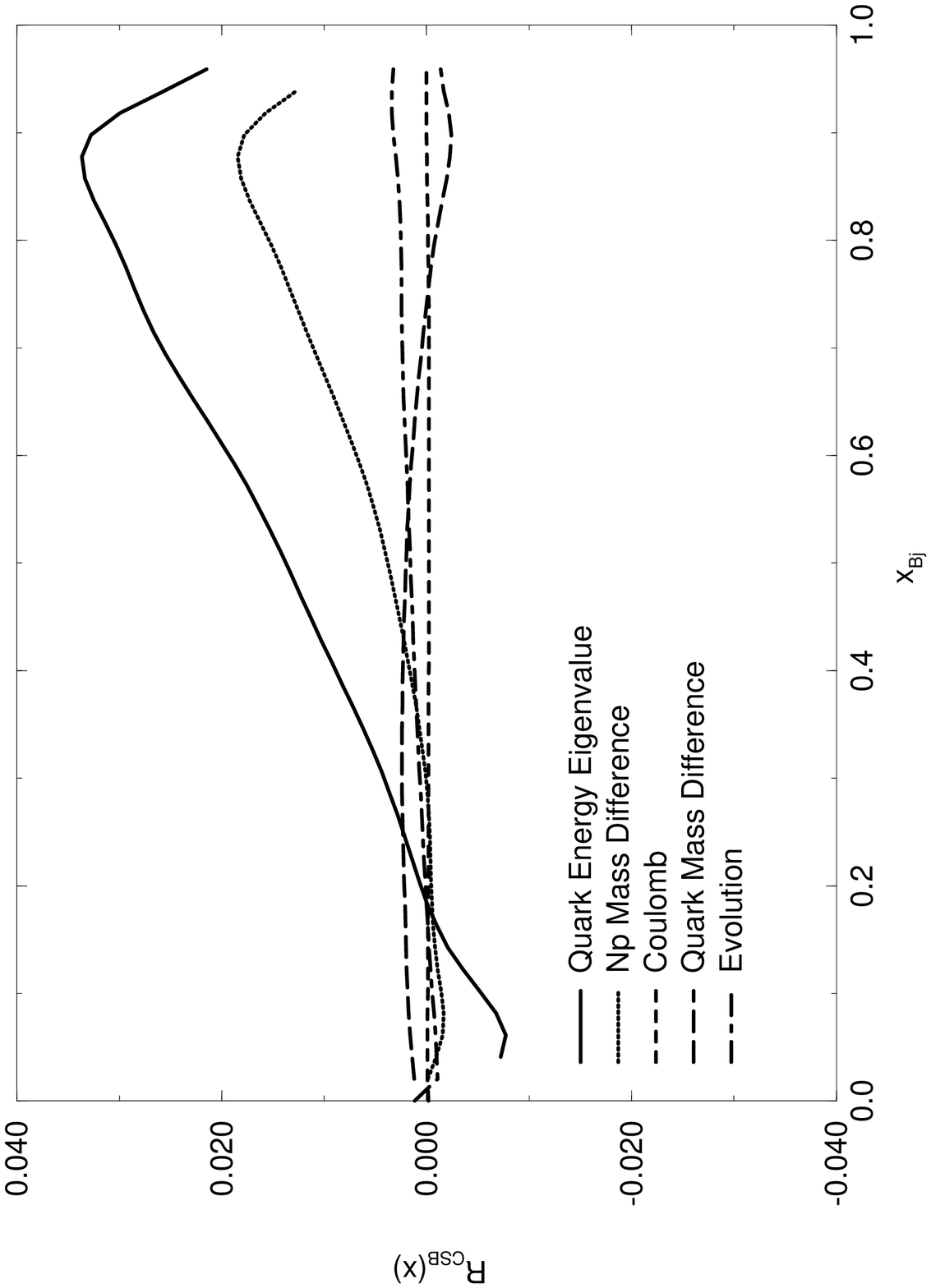}
\leftline{\hfill\vbox{\hrule width 5cm height0.001pt}\hfill}

\fcaption{ Minority quark ratio ,$R_{min}(x)$, for the LAMP at 10 GeV$^2$.}

\label{fig:minratio}

\end{figure}
   These results agree with the qualitative results obtained in Ref. 3,
where a model independent analysis of {\it some} of the corrections
discussed here was performed. They contrast markedly, however, with the
results of Ref. 4, where CSB effects on the
order 10-15\% were found for $x$ near 0.7. This huge result comes about
as a result of extreme sensitivity of the quark distribution to the
diquark mass parameter introduced in the Adelaide group`s method for
extracting valence distributions from quark models. The question of
which of these procedures is correct cannot be resolved at the level of
model building, but may be resolved experimentally in Drell-Yan
experiments\cite{3}. The result of these experiments will undoubtedly
give us new insight into the internal dynamics of the strong interaction.

\end{document}